\documentclass[useAMS, usegraphicx]{mn2e}
\usepackage{times}
\usepackage{amsmath}
\usepackage{rotating}
\long\def\symbolfootnote[#1]#2{\begingroup%
\def\thefootnote{\fnsymbol{footnote}}\footnote[#1]{#2}\endgroup}
\newcommand{\gae}{\lower 2pt \hbox{$\, \buildrel {\scriptstyle >}\over {\scriptstyle \sim}\,$}}
\newcommand{\lae}{\lower 2pt \hbox{$\, \buildrel {\scriptstyle <}\over {\scriptstyle \sim}\,$}}

\def\rmx{{\rm x}}

\begin{document}

\title[Maximum synchrotron frequency for shocks]{Maximum 
Synchrotron Frequency for Shock Accelerated Particles}

\author[Kumar, Hern\'andez, Bo\v snjak, Barniol Duran]{P. Kumar$^1$\thanks
{E-mail: pk@astro.as.utexas.edu, robertoh@physics.utexas.edu, zeljka.bosnjak@cea.fr, rbarniol@phys.huji.ac.il},
R. A. Hern\'andez$^2$\footnotemark[1], \v Z. Bo\v snjak$^3$\footnotemark[1] and R. Barniol Duran$^4$\footnotemark[1] \\
$^{1}$Department of Astronomy, The University of Texas, Austin, TX 78712, USA \\
$^2$Department of Physics, The University of Texas, Austin, TX 78712, USA \\
$^{3}$AIM (UMR 7158 CEA/DSM-CNRS-Universit\'e Paris Diderot) Irfu/Service
d'Astrophysique,  Saclay, 91191 Gif-sur-Yvette Cedex,  France\\
$^{4}$Racah Institute for Physics, Edmund J. Safra Campus, Hebrew University of Jerusalem, Jerusalem 91904, Israel}

\date{Accepted 2012 August 29. Received 2012 August 24; in original form 2012 January 10}

\pubyear{2012}

\maketitle

\begin{abstract}
It is widely believed that the maximum energy of synchrotron 
photons when electrons are accelerated in shocks via the Fermi
process is about 50 MeV (in plasma comoving frame). We show that
under certain conditions, which are expected to be realized in relativistic 
shocks of gamma-ray bursts, synchrotron photons of energy much larger 
than 50 MeV (comoving frame) can be produced. The requirement is that
magnetic field should decay downstream of the shock front on a length
scale that is small compared with the distance traveled by the highest
energy electrons before they lose half their energy; photons of energy much 
larger than 50 MeV are produced close to the shock front whereas 
the highest Lorentz factor that electrons can attain is controlled 
by the much weaker field that occupies most of the volume of the 
shocked plasma.
\end{abstract}


\section{Introduction}

High energy radiation from many astrophysical objects is believed
to be generated in shock heated plasma where charged particles are 
accelerated by the Fermi mechanism, eg. Fermi (1949), Axford et al. (1977), 
Bell (1978), Blandford \& Ostriker (1978), Achterberg et al. (2001), Sironi \&
Spitkovsky (2011). The maximum particle energy is set by the condition that 
radiative losses between acceleration episodes (which is the time it takes 
to travel from one side of shock front to the other) should 
be smaller than the energy gain. This upper bound to
electron energy means that the emergent synchrotron
radiation falls off to zero above a certain cut-off frequency
which is about 50 MeV in the plasma rest frame independent of 
the magnetic field strength in the shock heated fluid (de Jager \& Harding,
1992; Lyutikov, 2009; Piran \& Nakar, 2010; the derivation 
is provided in the next section). 

Many astrophysical objects, such as supernova (SN) remnants, pulsars,AGNs and gamma-ray bursts (GRBs), emit photons of energy larger than $\sim 10^2$MeV, and this upper limit has been used to 
eliminate the electron synchrotron process in shock-heated plasma 
as the radiation mechanism for these objects.
However, we show in this Letter that the $\sim 50$ MeV upper limit 
can be violated under certain astrophysically realistic conditions.

The upper limit of 50 MeV for synchrotron radiation is arrived 
at by assuming a uniform magnetic field throughout shock heated 
plasma. We show that this upper limit can be considerably
raised when the magnetic field is not uniform. A particular
case we analyze in some detail is where the magnetic field
decays with distance down-stream of the shock front, and 
find that for a range of parameters there can be a significant 
synchrotron flux even at several GeV (in fluid comoving frame).
This is described in section 2. The calculations 
presented in section 2 can be easily generalized to consider a 
situation where magnetic field fluctuates randomly instead of
decaying with distance from the shock front. The application to a few
astrophysical systems are presented in \S3.

\section{Maximum synchrotron frequency for shock heated plasma}

The argument that the maximum photon energy for synchrotron 
emission for shock accelerated electrons is $\sim 50$ MeV is
straightforward and goes as follows.

Electrons gain energy by a factor $\sim 2$ whenever they cross 
a relativistic shock front and are scattered back to the other side. The
time it takes for a charge particle of Lorentz factor $\gamma_e$
to make this trip from one side of the shock front to the
other is of order of the Larmor time, $t_L = m c \gamma_e/
(q B)$; where $B$ is the comoving frame magnetic field strength, 
$m$ is the particle mass, and $q$ is its charge. The energy lost to
synchrotron radiation during this time is $\delta E \sim t_L
  \sigma_T B^2 \gamma_e^2 c/6\pi \sim \sigma_T B \gamma_e^3 m c^2/
(6\pi q)$. Particle acceleration ceases when $\delta E\sim
  m c^2 \gamma_e/2$, and therefore the maximum Lorentz factor 
 a particle can attain is given by
\begin{equation}
    {\sigma_T B \gamma_e^2 \over 3 \pi q} \sim 1 \quad{\rm or}
    \quad   \gamma^2_{max} \sim {9 m^2 c^4\over 8 B q^3},
    \label{gam_max1}
\end{equation}
where $\sigma_T = 8\pi q^4/(3 m^2 c^4)$ is the Thomson 
cross-section. The synchrotron photon energy corresponding to
$\gamma_{max}$ is
\begin{equation}
   h\nu_{max} \sim {q B \gamma_{max}^2 h \over 2 \pi m c} \sim
    {9 m c^3 h \over 16\pi q^2}
\end{equation}
Thus, the maximum photon energy for 
electrons is $\sim 50$ MeV and for protons $\sim 10^2$ GeV\footnote{Although
proton synchrotron process can produce $\gamma$-rays with energy of
$\sim 10^2$GeV in plasma comoving frame, this is a very inefficient
mechanism and not likely to play a significant role in GRBs.}. These
numbers might be overestimated by a factor 5-10 due to uncertainty
regarding the time it takes particles to travel from one side 
of the shock front to the other (which could be an order of magnitude
larger than the Larmor time assumed in these calculations, e.g. Achterberg 
et al. 2001, Lemoine \& Revenu 2006, Sagi \& Nakar 2012, Uchiyama et al. 2007).
 All calculations in this paper are affected 
by this uncertainty. However, the ratio of $\nu_{max}$ for the 
case of a uniform magnetic field  to that for an inhomogeneous field 
geometry, discussed below, should be fairly secure.

\subsection{Inhomogeneous magnetic field and the maximum 
   synchrotron frequency}

Let us consider that the magnetic field decays with distance
down-stream of the shock front as 
\begin{equation}
   B(x) = B_w (x/L_p)^{-\eta} + B_0
   \label{Bx0}
\end{equation}
where $L_p$ is the length scale over which the field decays, $B_w$ \& $B_0$
are the strongest and weakest magnetic field strengths in the shocked fluid, 
and $x\ge L_p$.  If magnetic field is generated by the Weibel mechanism then
$L_p$ is of order the plasma length scale (Medvedev \& Loeb, 1999), i.e. 
\begin{equation}
   L_p = [m_p \Gamma_s c^2 /(4\pi n_e q^2)]^{1/2} = 
    2.2\rmx10^7 {\rm cm} (\Gamma_s/n_e)^{1/2}
  \label{Lp}
\end{equation}
where $m_p$ is proton mass, $\Gamma_s$ 
is the Lorentz factor  of the shock front wrt the unshocked
fluid, and $n_e$ is the number density of electrons in the shocked
fluid comoving frame\footnote{We note that for relativistic 
shocks there is little difference between proton and electron 
plasma length scales as long as electrons and protons are in
rough equipartition which is found to be the case from the
study of numerous gamma-ray burst afterglows, e.g. Panaitescu 
\& Kumar (2002).}. 

Particles accelerated in a shock influence the generation of magnetic
fields and can substantially increase the length scale for field decay
(Medvedev et al. 2005; Medvedev \& Zakutnyaya, 2009; Keshet et al. 2009).
A larger coherence length magnetic field produced by high energy 
particles -- which have larger plasma scale -- decays on a longer
time scale. However, these fields are also likely to 
be weaker (Medvedev \& Zakutnyaya, 2009), although their true strength is 
quite uncertain. We, therefore, use measurements of magnetic fields 
in GRB afterglows for guidance. The volume averaged magnetic field 
energy density in GRB external shocks, while they are relativistic,
is found to be a factor $\sim 10^4$ smaller than the equipartition 
value for a good fraction of bursts (Panaitescu 
\& Kumar, 2001, 2002; Bjornsson et al. 2004; Wei et al. 2006; 
Rykoff et al. 2006; Rol et al. 2007; Chandra et al. 2008; Liang et al. 2008; Gao et al. 2009;
Xu et al. 2009; Cenko et al. 2010; Rossi et al. 2011;
Kumar \& Barniol Duran, 2010). This
suggests that the field decays with distance downstream by a factor 
$\sim 10^2$ since the strength close to the shock front, due to
the Weibel instability, is of order of the equipartition value.
The length scale for this decay, however, is uncertain. For the sake 
of concreteness we take $L_p$ to be plasma scale for our calculations.
Fortunately, the results presented in this
paper don't depend on the precise length-scale for field decay as long
as it is much smaller than the distance traveled by the highest energy 
electrons before they lose their energy to radiation. 

The Larmor radius of an electron, $R_L(\gamma_e)=m_e c^2 \gamma_e/(q B)$, 
increases with distance from the shock front as the magnetic field gets 
weaker (eq. \ref{Bx0}), and an electron traveling down-stream
is likely to be sent back upstream when $R_L \lae x$ ($x$ is
the distance from the shock-front). Therefore, an
absolute upper limit on $\gamma_e$ is set by the requirement that
$R_L$ is smaller than the width of the shock heated plasma.

A more stringent upper limit on $\gamma_e$ is obtained by the
requirement that energy lost by an electron to radiation 
while traveling from up-stream to down-stream (in between
two consecutive episodes of energy gain) should be smaller than
$m_e \gamma_e c^2/2$.

The energy loss rate for an electron of LF $\gamma_e$
due to synchrotron radiation while traveling down-stream of the shock front is
\begin{equation}
   {d (m_e c^2 \gamma_e) \over dt} = - {\sigma_T\over 6\pi} 
   B^2 \gamma_e^2 c = - {\sigma_T \gamma_e^2 c\over 6\pi} \left[ 
    B_w \left( {L_p\over x}\right)^\eta + B_0\right]^2.
\end{equation}
We are interested in the case where $B_w\gg B_0$.
However, since $L_p$  -- the length scale over which the magnetic field 
decays -- is much-much smaller than the thickness of the shocked plasma,
most of the synchrotron loss occurs in the region of low magnetic field
($x\gg L_p$), which also controls the maximum Lorenz factor of electrons,
provided that
\begin{equation}
   B^2_w \left[ 1 - (L_p/x_0)^{2\eta-1} \right]/(2\eta-1) <
        B_0^2 \, (R_L/L_p),
\end{equation}
where $x_0/L_p\equiv (B_w/B_0)^{1/\eta}$; the above equation is obtained
by calculating energy loss in the region of low magnetic field ($B_0$) where 
the electron travels a distance $\sim R_L$, and the loss in the region 
$L_p\lae x<x_0$, and requiring the former to be larger.
Since $x_0/L_p\gg 1$ for the case of interest, the above condition 
simplifies to 
\begin{equation}
  (B_w/B_0)^2 \lae R_L/L_p
   \label{Bw_B0}
\end{equation}
 when $\eta>1/2$. Thus, one of 
the requirements for exceeding the $\sim 50$ MeV limit is that the width of the
region occupied by high magnetic fields ($x_0$) is much smaller 
than $R_L$. The above condition (eq. \ref{Bw_B0}) also guarantees that 
the Larmor radius of electron
in the region of high magnetic field is much larger than $L_p$, and thus 
the deflection of electron orbit while passing through this region is small;
electrons are turned around in the region of low magnetic field which occupies
 most of the shocked plasma volume.

The maximum LF of an electron, $\gamma_{max}$, is obtained by 
requiring that the energy loss due to synchrotron radiation 
in a Larmor time ($R_L/c$) not exceed $m_e\gamma_{max} c^2/2$.
The case of interest is where electrons lose their energy while 
traveling through the region of weak magnetic field. In this case 
$\gamma_{max}$ is same as given in equation (\ref{gam_max1}), i.e.
\begin{equation}
   \gamma_{max}^2 \sim {9 m^2 c^4\over 8 q^3 B_0}.
   \label{gam_max}
\end{equation}
This Lorentz factor must satisfy the condition that $R_L(\gamma_{max})$
is smaller than the comoving radial width of the shocked fluid (the width
is $R/\Gamma$ by causality argument), i.e. the electron is confined to
the system. This requires  
\begin{equation}
   B_0 \gae \left[ {3\pi \over \sigma_T q}\right]^{1/3} (m_e c^2
    \Gamma/R)^{2/3} = 0.1\,{\rm mG}\; (\Gamma/R_{17})^{2/3},
   \label{B0_min}
\end{equation}
where $\Gamma$ is the Lorentz factor of the shocked fluid, $R$ is the
distance of the shock front from the center of explosion, and we have
adopted the widely used convenient notation $x_n\equiv x/10^n$.

Electrons also suffer IC loss of energy which is ignored here;
considering that $\gamma_{max}\sim 10^8 B_0^{-1/2}$, IC
scatterings are in Klein-Nishina regime and thus the IC loss
is highly suppressed (Barniol Duran \& Kumar, 2011).

The maximum synchrotron frequency, in shock comoving frame, 
 is given by
\begin{equation}
   \nu_{max} \sim {q \gamma_{max}^2 B_w\over 2\pi m_e c} \sim
      {9 m_e c^3 \over 16\pi q^2} \left( {B_w\over B_0}\right)
      \sim 50\,{\rm MeV} (B_w/B_0),
   \label{numax_new}
\end{equation}       
which is larger than the case of a uniform magnetic field by a 
factor $B_w/B_0$.

We next estimate the synchrotron flux at $\nu_{max}$ to see whether
it lies on the powerlaw extension of flux at lower frequencies 
($< 50$ MeV) or not.

Let us consider electrons of LF $>\gamma_{max}/2^{1/2}$ which 
produce synchrotron photons of frequency $>\nu_{max}/2$.
By the definition of $\gamma_{max}$ these 
electrons will lose half of their energy while traveling down-stream of
the shock front. The energy fraction of these electrons lost to
synchrotron radiation ($\xi$) while they travel through the 
region where the magnetic field strength is $\sim B_w$ is 
\begin{equation}
    \xi \sim {B_w^2 L_p\over B_0^2 R_L(\gamma_{max})} \sim 
    2\rmx10^{-4} \left[ 
   {B_w\over B_0}\right]^2  B_0^{3/2} (\Gamma_s/n_e)^{1/2},
   \label{xi1}
\end{equation}
where $R_L(\gamma_{max}) = m c^2 \gamma_{max}/(q B_0)$ is the
Larmor radius, and we made use of equations (\ref{Lp}) \& 
(\ref{gam_max}) to substitute for $L_p$ \& $\gamma_{max}$.
The frequency of radiation produced in the region of high magnetic
field ($B_w$) is $\sim\nu_{max}$ which is given by equation 
(\ref{numax_new}).
Much of the rest of the electron energy is lost to radiation in 
the region occupied by the weaker field $B_0$, and the synchrotron 
frequency of the emergent radiation from this region is 
$\nu_{max} (B_0/B_w) \equiv \nu_{low}$;
where $\nu_{low}\sim 50$ MeV is the maximum synchrotron frequency
when the magnetic field is uniform. 

The specific flux at $\nu_{max}$ is obtained from the equation
$\nu_{max} f(\nu_{max}) \sim \xi \nu_{low} f(\nu_{low})$.
Therefore, the spectral index between $\nu_{low}$ and 
$\nu_{max}$ --- defined as $f_\nu\propto \nu^\beta$ --- is
\begin{equation}
   \beta = - 1 + { \ln\xi\over \ln(\nu_{max}/\nu_{low}) } = 
  1 + { 1.5\ln(B_0)+ 0.5 \ln(\Gamma_s/n_e) - 8.5 \over 
    \ln(B_w/B_0) }.
   \label{beta}
\end{equation}
The second equality is obtained by substituting for $\xi$ from
equation (\ref{xi1}).
It can be shown, after some algebra, that $\beta$ 
cannot be larger than $-p/2$, where $p$ is the index for 
electron distribution. 

We note that the synchrotron radiation formula can only be 
used provided that the magnetic field is coherent on
scale $\sim R_L/\gamma_{max} \sim 
m_e c^2/(q B_w)\sim 1.7\rmx10^3\, B_w^{-1}$cm. The coherence 
scale for magnetic field close to the shock-front is
$\sim L_p \gae 2\rmx10^7 (\Gamma_s/n_e)^{1/2}$ cm. Thus, we see
that the coherence length for magnetic fields is much larger than 
$R_L/\gamma_{max}$ as long as $B_w \gae $ 0.1 mG. The
coherence scale grows with increasing distance from the shock
front (since smaller scale fields have time 
to decay), and so does the Larmor radius. However, as long as
$B_0\not\ll 1$ mG, the magnetic field is coherent on length
scales that are much larger than $R_L/\gamma_{max}$ everywhere
down-stream of the shock front. Therefore, it is safe to use 
the synchrotron radiation results as we have in this
section\footnote{When magnetic field changes on a length
scale that is short compared with electron Larmor radius divided 
by its Lorentz factor, 
 the radiation produced is in the jitter regime which is discussed 
in Medvedev, (2000).}.

\section{Applications to GRBs and supernova remnants}

We consider two particular applications of the results found in the
previous section, one of which is the application to GRB afterglow radiation, and the other is the application to synchrotron radiation from SN remnants.

\subsection{Gamma-ray Bursts}

The relativistic jet produced in GRBs drives a shock wave into
the medium in the vicinity of the burst, and synchrotron radiation
from shock heated plasma is responsible for the long lived 
afterglow radiation from radio to $\gamma$-ray frequencies (eg. Piran, 2004; 
Meszaros 2006; Gehrels et al. 2009; Zhang 2007). Roughly half of the GRB 
jet energy is imparted to the surrounding medium at the deceleration 
radius, $R_d$, which is given by, eg. Piran (2004), Zhang (2007)
\begin{equation}
   R_d = \left[ {3 E\over 4\pi n_0 m_p c^2 \Gamma_0^2}\right]^{1/3}
        \sim 1.2\rmx10^{17}{\rm cm} (E_{53}/n_0)^{1/3} 
      \Gamma_{0,2}^{-2/3},
\end{equation}
where $E$ is the isotropic equivalent of energy carried by the jet,
$n_0$ is the average number density of protons in the unshocked medium 
within the region of radius $R_d$, and $\Gamma_0$ is the initial Lorentz 
factor of the jet. The Lorentz factor of the jet decreases with distance as 
$R^{-3/2}$. The comoving radial width of the shocked plasma is $\delta R\sim 
R/\Gamma$, and therefore $\delta R/L_p \sim 10^8 R_{17} \Gamma_{2}^{-1}
 (n_e/\Gamma)^{1/2}$.

The minimum magnetic field expected in the shocked fluid is
$B_0 = 4 \Gamma B_{ism}$ which is simply the field of the GRB 
circum-stellar-medium (CSM) compressed by the relativistic shock 
by a factor $4\Gamma$; where $B_{ism}$ is the field in the CSM at a distance 
$\sim R_d$ from the center of explosion and is expected to be of order
a few $\mu$G.
Thus, the condition that electrons of Lorentz factor $\gamma_{max}$ are
confined to the shocked fluid by the weak magnetic field (see 
eq. \ref{B0_min}) is satisfied for GRB external shocks as long as 
the shock is highly relativistic ($\Gamma\gae 10$).

The field is likely amplified by a large factor immediately 
down-stream of the shock front by the two-stream instability, 
such that the energy density in the magnetic field is a fraction, 
$\epsilon_B$, of the thermal energy density of the shocked 
fluid (Medvedev \& Loeb, 1999). In this case
\begin{equation}
   B_w \sim (32 \pi \epsilon_B n_0 m_p c^2 \Gamma^2)^{1/2},
\end{equation}
with $\epsilon_B\sim 10^{-1}$ (Medvedev \& Loeb, 1999); we have used here
the relativistic self-similar shock solution (Blandford \& McKee, 1976)
according to which the thermal energy density behind the shock
front is $4 n_0 m_p c^2 \Gamma^2$. This magnetic field has 
a coherence length of order the plasma scale and decays down-stream 
of the shock front (Gruzinov, 2001; Sironi \& Spitkovsky, 2011; 
but see also Medvedev et al. 2005).  It is unclear what is
the exact value of the field very far from the shock front. 
According to recent numerical simulations of Sironi \& Spitkovsky (2011) 
the far field might be of order the shocked compressed ISM field;
empirically we know from GRB afterglows that $\epsilon_B\lae10^{-4}$ for
a large fraction of bursts (see \S2.1 for references to a number of papers
on this topic). Therefore, for GRB afterglows we expect the ratio between
$B_w$ and $B_0$ to be larger than $\sim10^2$ and possibly of order
\begin{equation}
    {B_w\over B_0} \sim {(2\pi \epsilon_B n_0 m_p)^{1/2} c\over
     B_{ism} } \sim 9.5\rmx10^3 \, (\epsilon_B n_0)^{1/2}
    B_{ism,-5}^{-1},
   \label{Bw_B0}
\end{equation}
where $B_{ism,-5}$ is the ISM magnetic field in units of 10 $\mu$G;
the precise value of $B_w/B_0$ is not important as long as it is
larger than $\sim 10$, and the field decays on a length-scale
that is short compared with the Larmor radius ($R_L$) or the distance the
highest energy electrons travel before losing half their energy.
Substituting this into equation (\ref{numax_new}) we find
\begin{equation}
  h\nu_{max} \sim 0.5\, {\rm TeV} \, (\epsilon_B n_0)^{1/2}
     B_{ism,-5}^{-1}.
    \label{numax3}
\end{equation}
The maximum energy for synchrotron photons in the observer frame
is $\Gamma\nu_{max}/(1+z)$. If the time it takes for an electron
to travel from one side of the shock front to another is larger than
$R_L/c$ by a factor 5--10 then $\nu_{max}$ would be smaller than given
by the above equation by the same factor.

\begin{figure}
 \includegraphics[width=\linewidth]{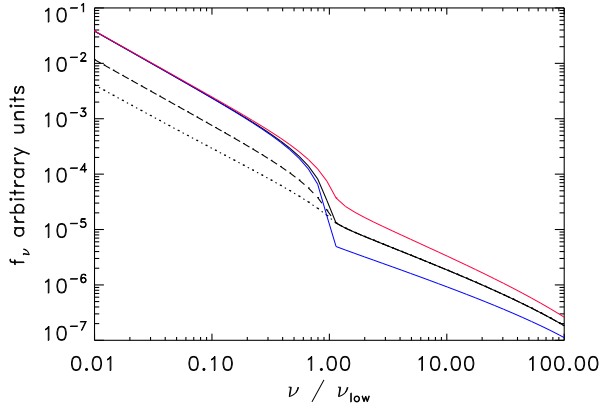}
 \caption{Numerical spectra for different values of $R_L/L_p$ and $\eta$. 
The solid, dashed and dotted lines correspond to $R_L/Lp = 10^7$, $3 \times 
10^6$ and $10^6$, respectively (and $\eta = 0.7$). Similarly, $R_L/Lp = 10^7$ 
and $\eta = 0.6$, $0.7$ and $0.9$ for the red, black and blue lines, 
respectively. For all these cases, $\gamma_{\text{max}}/\gamma_{\text{min}} = 
10^5$ and $B_w/B_0 = 10^3$, and the size of the system is larger than $R_L$ 
by a factor of $\sim10$. The spectrum above $\nu_{low}\equiv 9 m_e c^3/(16\pi 
q^2)\sim 50$ MeV, has a power-law slope roughly equal to the spectrum at lower 
frequencies and it extends to well beyond $\nu_{low}$. The flux at 
$\nu_{low}$ can exhibit a sharp decline by almost an order of magnitude for 
certain parameters, however, the flux at $\gae 1$GeV even in the worst case 
lies on a slightly steeper power-law extension of the lower energy 
($\lae 50$ MeV) spectrum.  }
 \label{spectra}
\end{figure}

Using equations (\ref{beta}) \& (\ref{Bw_B0}) one can obtain an approximate
analytical power-law spectral index for $\nu_{low} < \nu < \nu_{max}$;
however, we have calculated the spectrum numerically for different choices of parameters and present the results
in Figure ~\ref{spectra}.

The spectrum in the observer frame should extend to $\sim$ TeV 
energies provided that IC losses don't prevent electrons from 
accelerating to $\gamma_{max}\sim10^8$ (eq. \ref{gam_max}), and these high 
energy photons are not converted to pairs while traveling from the emission 
site to us\footnote{Photons of energy $\gae$TeV from a source at cosmological
distance are converted to electron-position pairs due to interaction with
infrared background radiation.}; the effect of IC loss on $\gamma_{max}$,
in the context of GRBs, is discussed in Barniol Duran \& Kumar (2011)
and Piran \& Nakar (2010). For certain range of parameters the flux at 
photon energies much larger than 50 MeV$\times \Gamma$ lies below the 
powerlaw extrapolation of flux below $\nu_{low}$ (Fig. 1), and in those 
cases the flux at very high energies is likely to be lower than the 
sensitivity limit of current detectors.

Several bursts detected by the Fermi satellite show emission
above 10 GeV in the GRB host galaxy rest frame: Abdo et al. (2009a,b, 2010), 
Ackermann et al. (2010, 2011). These 
bursts violate the conventional upper limit of $\sim 50 \Gamma/(1+z)$
 MeV $\sim5$ GeV for synchrotron radiation from shock heated plasma, 
eg. Piran \& Nakar (2010). However, the $>$10$^2$ MeV data for
these bursts is in excellent agreement with 
synchrotron radiation from the external shock, which also provides 
a very nice fit to the late time x-ray, optical and radio data
using the same exact parameters that provide a fit to the early
$>$10$^2$ MeV data (Kumar \& Barniol Duran, 2009, 2010). 

This conflict between the excellent fit to the high energy data on 
one hand and the conventional upper limit on synchrotron photon energy 
of $\sim$ 5 GeV on the other hand can be resolved provided that the magnetic 
field just behind the shock front is much larger than far down-stream;
photons of energy $\gae 5$ GeV are produced by electrons of LF
$\sim\gamma_{max}$ while they travel through regions of stronger magnetic 
field close to the shock front. 

We note that there is a large uncertainty regarding the length scale
over which fields decay, $B_w$ \& $B_0$. However, the comoving frame 50 MeV
limit on synchrotron photons is violated whenever $B_w/B_0\gae 10$, and
the distance electrons of LF $\gamma_{max}$ travel before losing their
energy to radiation is large compared to the length-scale for the
decay of magnetic field. The data for GRBs 
with $\gae 10$ GeV emission show that 
the volume averaged magnetic field in the shocked fluid is a factor 10$^3$
smaller than the equipartition value\footnote{The small value for volume-averaged magnetic field for these GRBs follows from the fact that the 
synchrotron cooling frequency at the end of the burst must have been 
$\gae 10$ MeV; otherwise the ratio of flux at 150 keV to that at 100 MeV would 
be much larger than the observed value (large cooling frequency, 
$\gae 10$ MeV, implies small magnetic field).} (Kumar \& Barniol Duran, 2010) 
whereas the Weibel generated field strength in the immediate vicinity of the
shock front is expected to be $\sim30$\% of the equipartition value
(Medvedev \& Loeb, 1999), and therefore $B_w/B_0\gae 10^2$ for these bursts.

\subsection{Supernova remnants}

The deceleration radius for supernova remnants is 
\begin{equation} 
   R_d \approx \left[ {3 E \over 4\pi n_0 m_p v_0^2}\right]^{1/3}
    \approx 5\rmx10^{18} (E_{51}/n_0)^{1/3} v_{0,9}^{-2/3}\,{\rm cm},
\end{equation}
where $E$ is the kinetic energy of the remnant, and $v_{0,9}$ is its
initial speed in unit of 10$^9$ cm/s. The radius increases with time as 
$t^{2/5}$ during the adiabatic expansion phase of the remnant. The radial 
width of the remnant ($\delta R$) is roughly 1/5 of its radius, and the 
proton-plasma length $L_p \sim 2\rmx10^7 n_0^{-1/2}$ cm. Therefore, 
$\delta R/L_p \gae 10^{11}$.

The equipartition magnetic field for the remnant is
\begin{equation}
  B_w\sim (40 \pi m_p n_0)^{1/2} v \sim 1\, {\rm mG}\, n_0^{1/2} v_8,
\end{equation}
where $v$ is the remnant speed. The mean magnetic field of the
remnant should be larger than, or equal to, the shock compressed
magnetic field of the CSM, i.e. $B_0 \gae 5 B_{ism}\sim 10\mu$G.
Therefore, $B_w/B_0\lae 10^2$.

The maximum Lorentz factor of electron for this mean magnetic
field is $\gamma_{max}\sim [3\pi q/(\sigma_T B_0)]^{1/2} \sim
3\rmx10^{10}$ (see eq. \ref{gam_max}), and its Larmor radius is
5x10$^{18}$cm which is smaller than the shell width when the velocity
drops below 10$^8$cm/s.

The synchrotron cooling time for electrons with Lorentz factor
 $\gamma_{max}\sim 10^{10}$ traveling in a field of strength $B_0$,
which occupies much of the volume of the remnant, is $t_c \sim 
6\pi m_e c/(\sigma_T B_0^2 \gamma_{max})$. In other words, the distance
electrons travel before cooling down is $L_c \sim c t_c \sim 
  8\rmx10^{18}$ cm which is of order of the remnant width when the
remnant velocity drops below 10$^8$cms$^{-1}$.

Thus, $L_c/L_p\sim 10^{11}$, and so $L_c/L_p\gg (B_w/B_0)^2\sim 10^4$.
Therefore, the specific flux at 50 MeV $(B_w/B_0)\sim 5$ GeV is a factor
$\xi\sim 10^7$ smaller than the flux at 50 MeV, which is perhaps too 
small to be of practical consequence. However, if the magnetic field
in the remnant decays not on the plasma length scale but on $\sim 10^4$
times the plasma scale then the flux at $\nu_{max}\sim 5$ GeV would 
lie on the extension of the powerlaw spectrum at lower energies.

\section{Conclusions}

We have shown that photons of $\gae 10$ GeV energy observed from 
several GRBs by the Fermi satellite can be produced via the
synchrotron process in the shock heated circum-stellar medium.
The conventional wisdom that the maximum energy of photons 
in this situation should not exceed $\sim 50\Gamma/(1+z)$ MeV
$\sim 5$ GeV is violated because this limit is obtained by assuming a 
uniform magnetic field down-stream of the shock front.
However, when the field is much stronger close to the shock front
and decays downstream (such as when magnetic fields are generated
by Weibel-type instabilities\footnote{Large variations in magnetic
field strength across the shocked plasma could also arise as a result
of field in the circum-stellar medium varying with distance from the
progenitor star which is compressed and amplified near
the shock front. Long-duration GRBs and SNe are produced when a massive
star dies, and the medium within a few parsecs of the progenitor star is 
carved out by its wind within the last $\sim10^2$ years of the
collapse. Although little is known about the wind and its associated
magnetic field in the last $\sim10^2$years of a star's life, it is possible
that the magnetic field in the wind could undergo large variation
with time due to the magnetic cycle of the star driven by its rapid
rotation and sub-surface convection.}) the maximum photon energy is
larger than this limit by a factor of the ratio of 
field just behind the shock front and far downstream; photons
of energy $\gae 10^2$ GeV (in observer frame) can be generated 
this way in GRB early afterglows.

The upper limit of $\sim 50$ MeV is difficult to violate for synchrotron 
radiation from supernova remnants unless the field were to decay on a length
scale much larger than proton plasma length but smaller than the distance 
highest energy electrons travel before losing half their energy.

\section*{Acknowledgments}

This work has been funded in part by NSF grant 
ast-0909110, and a Fermi-GI grant (NNX11AP97G). ZB acknowledges 
the French Space Agency (CNES) for financial support.

\end{document}